\documentstyle[12pt,psfig]{article}
\begin{document}
\baselineskip 18pt
\def\kongetsu{\ifcase\month\or
 January\or February\or March\or April\or May\or June\or
 July\or August\or September\or October\or November\or December\fi
 \space\number\year}
\def\thebibliography#1{\section*{References\markboth
 {References}{References}}\list
 {[\arabic{enumi}]}{\settowidth\labelwidth{[#1]}
 \leftmargin\labelwidth
 \advance\leftmargin\labelsep
 \usecounter{enumi}}
 \def\newblock{\hskip .11em plus .33em minus .07em}
 \sloppy
 \sfcode`\.=1000\relax}
\let\endthebibliography=\endlist

\def\disp{\displaystyle}
\def\gsim{{\mathop >\limits_\sim}}
\def\lsim{{\mathop <\limits_\sim}}
\def\bbbar{$B^0$-$\bar{B}^0$}
\def\kkbar{$K^0$-$\bar{K}^0$}
\def\Dm{\Delta m}
\def\DM{\Delta M}
\def\r2{\sqrt 2}
\def\ra{r_\alpha}
\def\rb{r_\beta}
\def\rmuu{\gamma^{\mu}}
\def\rmud{\gamma_{\mu}}
\def\PL{{1-\gamma_5\over 2}}
\def\PR{{1+\gamma_5\over 2}}
\def\sw2{\sin^2\theta_W}
\def\v#1{v_#1}
\def\tb{\tan\beta}
\def\sb{\sin\beta}
\def\cb{\cos\beta}
\def\s2b{\sin 2\beta}
\def\c2b{\cos 2\beta}
\def\s2b2{\sin^22\beta}
\def\uL{{\tilde u}_L}
\def\uR{{\tilde u}_R}
\def\cL{{\tilde c}_L}
\def\cR{{\tilde c}_R}
\def\tL{{\tilde t}_L}
\def\tR{{\tilde t}_R}
\def\tf{{\tilde t}_1}
\def\ts{{\tilde t}_2}
\def\MCH{M_{H\pm}}
\def\mgr{m_{3/2}}
\def\m#1{{\tilde m}_#1}
\def\mH{m_H}
\def\mwi{{\tilde m}_{\omega i}}
\def\mw#1{{\tilde m}_{\omega #1}}
\def\mQ{{\tilde m}_Q}
\def\mU{{\tilde m}_U}
\def\stopmass{m_{\tilde t}}
\def\MuSL{m_{{\tilde u}L}^2}
\def\MuSR{m_{{\tilde u}R}^2}
\def\McSL{m_{{\tilde c}L}^2}
\def\McSR{m_{{\tilde c}R}^2}
\def\MtSk{m_{{\tilde t}k}^2}
\def\MtS#1{m_{{\tilde t}#1}^2}
\def\MM{M_{\tilde t}^2}
%

\begin{titlepage}
\hfill\vbox{\hbox{TKU-HEP 95/02}}

\hfill\vbox{\hbox{OCHA-PP-62}}

\hfill\vbox{\hbox{ \kongetsu}}
\vspace{2cm}

\begin{center}
{\Large
Constraints on Light
Top Squark  \\
from $B^0$-$\bar{B}^0$
mixing
}
\\
\vspace{2cm}
\renewcommand{\thefootnote}
{\fnsymbol{footnote}}
G. C. Cho $^1$
\footnote{Research Fellow of the Japan Society
for the Promotion of Science.},
Y. Kizukuri $^1$
\footnote{Supported by Grant-in-Aid for Scientific
Research from the Ministry of Education, Science
and Culture, contract number 6640409.}
\footnote{Deceased.},
and N. Oshimo$^2$\\
\vspace{0.5 cm}
{$^1$\it Department of Physics, Tokai University}\\
{\it 1117 Kita-Kaname, Hiratsuka 259-12, Japan}\\
{$^2$\it Department of Physics, Ochanomizu University}\\
{\it Otsuka 2-1-1, Bunkyo-ku, Tokyo 112, Japan}\\
\end{center}
\vspace{0.5cm}

\begin{center}
{\bf Abstract}
\end{center}
\vspace{0.6cm}

We discuss the constraints on the mass of the
lighter top squark from \bbbar\ mixing
in the minimal supersymmetric standard model.
A light top squark whose mass is less than half
of the $Z^0$-boson mass has not yet been
excluded from direct search experiments at LEP.
However,
the existence of the
light top squark may exceedingly enhance \bbbar\
mixing,
owing to the box diagrams exchanging the
charginos and the up-type squarks.
We show that for a sizable region of parameter space
the light top squark contribution to
\bbbar\ mixing becomes the same order of magnitude as
the standard $W$-boson contribution.  Taking into
account the experimental results for \bbbar\
and \kkbar\ mixings,
the existence of the light top squark
is excluded in an appreciable region of the
parameter space which LEP experiments have
not ruled out.
\end{titlepage}
\newpage
     Supersymmetric particles have been searched for
extensively in experiments, since
supersymmetry (SUSY) is considered one of
the most promising ideas for physics beyond the
standard model (SM) from both aesthetical and
phenomenological viewpoints.
Up to now no direct experimental evidence
for those new particles has been found,
giving constraints on SUSY parameters \cite{PDP}.
In particular, abundant data of $e^+e^-$ collisions at
LEP almost exclude pair productions of
charged SUSY particles at the $Z^0$ resonance.
However, there still remains a possibility that one of
the top squarks (hereafter,
we say ``stop" for short) has a mass less than the half
of the $Z^0$-boson mass.

     In the minimal supersymmetric standard model (MSSM)
one stop could naturally be lighter than the other
squarks \cite{ellis}.
Because, the large mass of the top quark induces a
large mixing between the left- and the right-handed stops,
while such a mixing is negligible in the first
and second generations.
One of the stops could be the lightest charged SUSY
particle.
Moreover, if the left-right mixing angle of the stops
has a certain value,
the lighter stop decouples with the $Z^0$-boson \cite{hikasa}.
Then the cross section of the stop pair production
is not large at the $Z^0$ resonance,
and the stop could escape detection at LEP.

     The experimental mass bounds on the lighter stop
$\tilde{t}_1$ have been given by CELLO at PETRA \cite{cello},
AMY, TOPAZ, VENUS at TRISTAN \cite{venus},
and OPAL at LEP \cite{opal}, which
depend on the assumed mass difference $\Dm$ between the stop
and the lightest neutralino.  Assuming $\Dm\geq 5$ GeV,
the lower bound is about 45 GeV, if the left-right mixing
angle has a value not close to the `decoupling' value.
On the other hand, if this mixing angle lies in the
vicinity of the `decoupling' value, the lower bound is
about 40 GeV for $\Dm\geq 5$ GeV and 25 GeV for
$\Dm\geq 2$ GeV.  There also have appeared several
analyses which constrain the light stop mass from
other phenomena, such as the decay
$b\rightarrow s\gamma$ \cite{okada} and
the cosmological mass density \cite{fukugita}.
However, the allowed region derived from
the $e^+e^-$ experiments is not much altered.

     In this letter we discuss the possibility of the
existence of the light stop, whose mass is smaller
than $\frac{1}{2}M_Z$, by investigating its effects on
\bbbar\ and \kkbar\ mixings.
In the MSSM, these mixings receive contributions from the
box diagrams mediated by the charginos and
up-type squarks and those mediated by the charged Higgs
bosons and up-type quarks, as well as the standard
box diagrams.
It was recently shown \cite{mix} that
these new contributions are comparable to the SM
contributions in sizable regions of the SUSY
parameter space.  In particular,
the chargino contributions generally
become large if one stop is much lighter
than the other up-type squarks, owing to less
efficient cancellation among different
squark contributions.
On the other hand, such theoretical predictions for \bbbar\
and \kkbar\ mixings can be well examined by
experiments available at present or in the near future
\cite{bb2},
which may constrain or reveal
the new contributions by the MSSM.  The aim of this paper is
to study the new contributions, concentrating on the
case where a light stop exists and decouples with
the $Z^0$-boson.
We will show, as a conclusion,
that within the SUSY parameter space
consistent with the light stop, there are wide  regions
which are not consistent with  the present experimental
data for \bbbar\ and \kkbar\ mixings
or will be explored  by future experiments at $B$-factories.

     First, we briefly review the masses and mixings of the
squarks \cite{oshimo}, assuming $N=1$ supergravity and
grand unification \cite{sugra}.
At the electroweak scale
the interaction eigenstate squarks are mixed in
generation space
through their mass terms.  For the up-type squarks
these generation
mixings in the mass-squared matrix are approximately
removed by the same matrices that diagonalize the mass
matrix of the up-type quarks.  Consequently,
the `super' Cabibbo-Kobayashi-Maskawa (CKM)
matrix, which describes the generation
mixings in the interactions of the down-type quark,
up-type squark, and chargino in mass eigenstate basis,
is the same as the CKM matrix of the quarks.

     In each flavor the left-handed and right-handed
squarks are mixed by the Yukawa interaction,
which is proportional to the corresponding quark mass.
 From the smallness of the $u$- and $c$-quark masses,
these mixings can be safely neglected
for the first two generations.
The masses of the left-handed squarks $\uL, \cL$ and the
right-handed squarks $\uR, \cR$ are given by
\begin{eqnarray}
\MuSL &=& \McSL
                  = \tilde{m}^2_Q + \cos 2\beta
                    (\disp{\frac{1}{2} - \frac{2}{3}
                     \sin^2 \theta_W }
                               ) M^2_Z, \nonumber \\
\label{sqmass}
\MuSR &=& \McSR
                  = \tilde{m}^2_U + \disp{\frac{2}{3}}
                    \cos 2\beta \sin^2 \theta_W M^2_Z, \\
     & & \tan \beta = \disp{\frac{v_2}{v_1}}, \nonumber
\end{eqnarray}
where $\v1$ and $\v2$ stand for the vacuum expectation values
of the Higgs bosons with the hypercharges $-\frac{1}{2}$ and
$\frac{1}{2}$, respectively.
The mass parameters $\mQ$ and $\mU$ are
determined by the gravitino mass $\mgr$ and the gaugino masses,
and $\mQ\simeq\mU\sim\mgr$.
For the third generation,
an appreciable mixing between $\tL$ and $\tR$
is induced by the large top quark mass $m_t$.
The mass-squared matrix for the stops is given by
\begin{eqnarray}
M_{\tilde t}^2 = \left(
\begin{array}{cc}
\MuSL +(1-|c|)m_t^2 & (\cot\beta\mH +a^*\mgr)m_t \\
\noalign{\vskip0.2cm}
(\cot\beta\mH^* +a\mgr)m_t &  \MuSR +(1-2|c|)m_t^2
\end{array}
\right),
\end{eqnarray}
where $\mH$ denotes the
higgsino mass parameter.
If the SU(2)$\times$U(1) symmetry is broken through
radiative corrections, the magnitude of $\mH$ is at
most of order of $\mgr$, and $\tb\ \gsim\ 1$.  The dimensionless
constants $a$ and $c$ depend on other SUSY
parameters:  $a$ is related to the breaking of local
supersymmetry and its absolute value is of order of unity;
$c$ is related to radiative corrections to the
squark masses and $|c|=0.1-1$.
The mass eigenstates of the stops $\tf$, $\ts$ are obtained by
diagonalizing the matrix $M_{\tilde t}^2$ as
\begin{eqnarray}
S_t M^2_{\tilde{t}} S_t^\dagger
      &=& {\rm diag}(\MtS1, \MtS2) \ \ (\MtS1 < \MtS2).
\end{eqnarray}
The unitary matrix $S_t$ and the stop masses
$m_{{\tilde t}i}$ are given by
\begin{eqnarray}
S_t &=& \left(
\begin{array}{rr}
\cos \theta_t & -\sin \theta_t \\
\sin \theta_t & \cos \theta_t
\end{array}
\right),                  \nonumber  \\
\tan \theta_t
&=& \disp{\frac{2(\MM)_{12}}
{-(\MM)_{11}+(\MM)_{22}+\sqrt{D}}}, \\
   m_{{\tilde t}1(2)}^2 &=&
    \frac{1}{2}\{(\MM)_{11}+(\MM)_{22} -(+)\sqrt{D}\},
                       \nonumber  \\
   & & D=\{(\MM)_{11}-(\MM)_{22}\}^2+4\{(\MM)_{12}\}^2,
                              \nonumber
\end{eqnarray}
where SUSY parameters are assumed to have real values.

     Let us consider the conditions which SUSY parameters
should satisfy in order for a stop to have a small mass
and decouple with the $Z^0$-boson.  The interaction Lagrangian
for the lighter stop and the $Z^0$-boson is given by
\begin{equation}
{\cal L} = -i\frac{e}{\sin 2\theta_W}(\cos^2\theta_t-\frac{4}{3}
     \sin^2\theta_W)Z_\mu\tf^*\partial^\mu\tf +{\rm h.c.},
\end{equation}
which vanishes for $\cos^2\theta_t=\frac{4}{3}\sin^2\theta_W$.
If the mixing angle $\theta_t$ has this `decoupling' value,
SUSY parameters satisfy the following equations:
\begin{eqnarray}
\mQ^2 &=& \left( -1+\frac{2-3r}{1-2r}|c|\right) m_t^2+\MtS1, \\
\mH &=& \tb\{-a\mgr\pm\sqrt{r(1-r)}\left(\frac{1}{2}\c2b
        \frac{M_Z^2}{m_t^2}+\frac{|c|}{1-2r}\right)m_t\},
\end{eqnarray}
where $r=\frac{4}{3}\sin^2\theta_W$ and we have put
$\mQ^2=\mU^2$.  The value of $\mQ^2$
is determined by $|c|$ and ${\stopmass}_1$.
The experimental lower bound on the squark masses of the
first two generations is about 150 GeV \cite{CDF},
which imposes the constraint $\mQ(\simeq\mU)>150$ GeV,
as seen from eq. (\ref{sqmass}).
For ${\stopmass}_1<45$ GeV,
this constraint leads to $|c|>0.6$,
whereas $|c|$ is theoretically at most unity, corresponding
to $\mQ\simeq 230$ GeV.  The squarks of the first two generations
are thus predicted to have masses not much larger than 200 GeV.
The value of $\mH$ depends on $\tb$ and $a\mgr$ as well as
$|c|$.
However, since the value of $a\mgr$ should not be
much different from $\mQ$,
only $\tb$ is an independent parameter.
In the MSSM, the lightest neutralino $\chi_1$
has to be lighter than the light stop, which gives
another condition for $\tb$, $\mH$, and the SU(2) gaugino
mass $\m2$.
Therefore, if there exists a light
stop which decouples with the $Z^0$-boson, the values of other
SUSY parameters are restricted within narrow ranges.

     The interactions of the chargino, up-type squark, and
down-type quark and those of the charged Higgs boson,
up-type quark, and down-type quark can add sizable new
contributions to \bbbar\ and \kkbar\ mixings through
box diagrams.
These short distance contributions are
experimentally measured by the mixing parameter
$x_d$ for \bbbar\ mixing and the $CP$ violation parameter
$\epsilon$ for \kkbar\ mixing.  Neglecting small new
contributions, these parameters are theoretically given by
\begin{eqnarray}
  x_d &=& {G_F^2\over 6\pi^2}M_W^2{M_B\over \Gamma_B}
    f_B^2B_B|V_{31}^*V_{33}|^2\eta_B|A_{tt}^W+A^C+A_{tt}^H|,
\label{xd}        \\
  \epsilon &=& -{\rm e}^{i\pi/4}
     {G_F^2\over 12\r2\pi^2}M_W^2{M_K\over \Delta M_K}f_K^2B_K
      {\rm Im}[(V_{31}^*V_{32})^2\eta_{K33}(A_{tt}^W+A^C+A_{tt}^H)
                                        \nonumber \\
       & &+ (V_{21}^*V_{22})^2\eta_{K22}A_{cc}^W
 + 2V_{31}^*V_{32}V_{21}^*V_{22}\eta_{K32}A_{tc}^W],
\label{ep}
\end{eqnarray}
where $f_B$, $B_B$, and $\eta_B$ represent
the decay constant, the bag factor, and the QCD correction
factor,
respectively, for the $B$-meson, and $f_K$, $B_K$, and
$\eta_{Kab}$ represent those for the $K$-meson.
The CKM matrix is denoted by $V$.  The contributions
of the $W$-boson, chargino,
and charged Higgs boson box diagrams are respectively
expressed as $A^W$'s, $A^C$, and $A^H_{tt}$, which
are explicitly given in refs. \cite{bb2,inami}.
In the SM, $x_d$ and $\epsilon$
are given by eqs. (\ref{xd}) and (\ref{ep}) with $A^C=A_{tt}^H=0$.
The term proportional to $(V_{31}^*V_{32})^2$ in eq. (\ref{ep})
is enhanced by the same amount as $x_d$.
The difference between the contributions of the MSSM and
the SM can thus be measured by the ratio
\begin{equation}
R={A_{tt}^W+A^C+A_{tt}^H\over A_{tt}^W}.
\label{Ratio}
\end{equation}
If new contributions are negligible, $R$ becomes unity.

\begin{figure}[t]
\begin{center}
\psfig{file=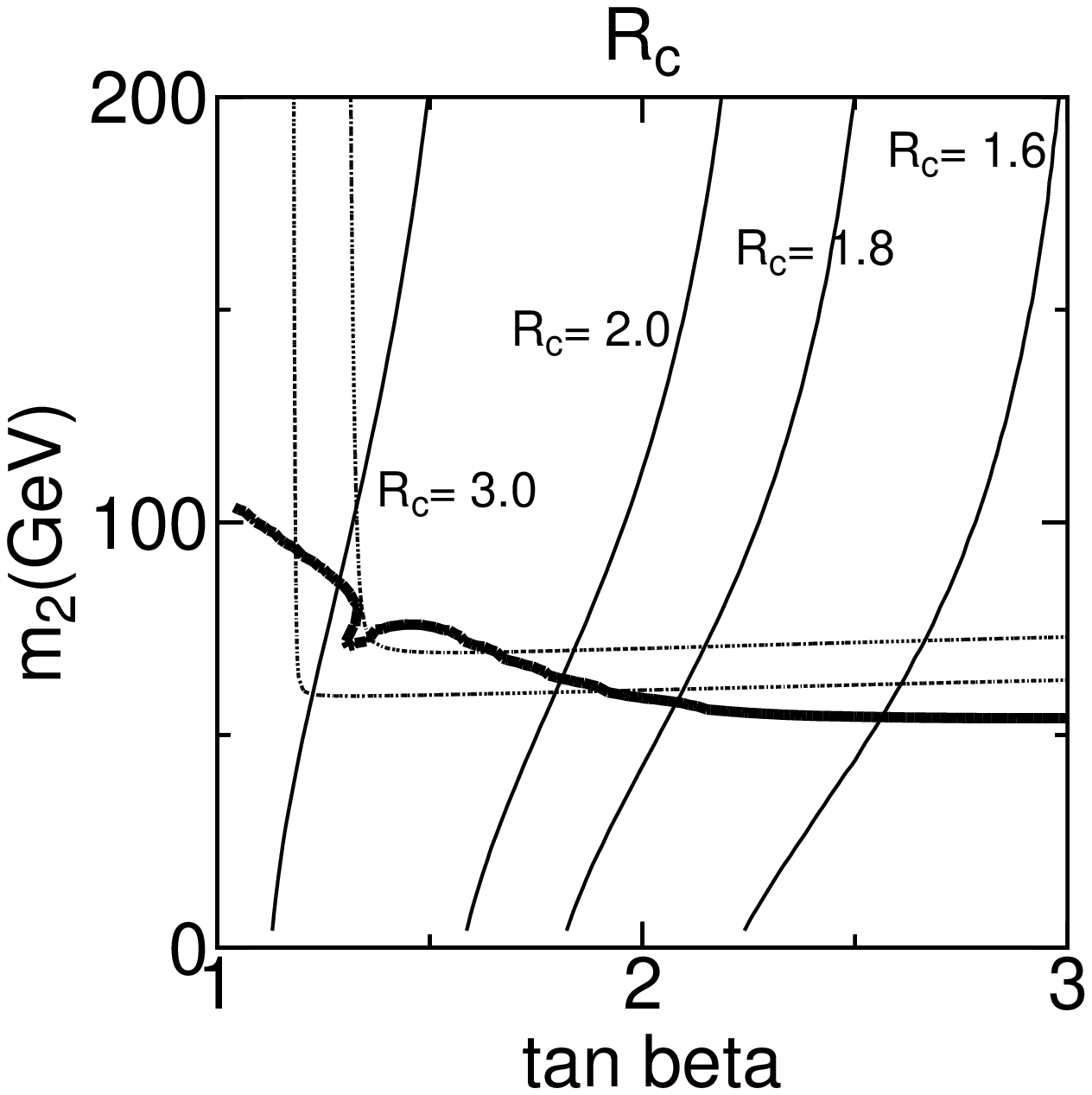,height=8cm}
\caption{The contours of the ratio $R_C$ for
${\stopmass}_1=40$ GeV
and $\cos^2\theta_t=\frac{4}{3}\sw2$.}
\end{center}
\end{figure}
     Now, we examine the effects of the light stop on
\bbbar\ and \kkbar\ mixings.
In order to see the chargino
contribution exclusively, we first consider the ratio
$R_C=(A_{tt}^W+A^C)/A_{tt}^W$ instead of $R$.
In Fig. 1 we plot contours of $R_C$
in the ($\tb, \m2$) plane, taking
$\cos^2\theta_t=\frac{4}{3}\sw2$
and ${\stopmass}_1=40$ GeV.
The values of the other parameters are set
for $|c|=0.7$ and $\mQ=\mU=a\mgr$.
For the $t$-quark mass we use $m_t=170$ GeV \cite{topmass}.
When the input parameters are fixed, two values are possible
for $\mH$ from eq. (7).  For $\tb=2$, the value of $\mH$ becomes about
$-70$ GeV or $-620$ GeV, which varies roughly in
proportion to
$\tb$.  We have taken the value of smaller magnitude.
The magnitude of the other one would be too large,
compared to $\mgr$, for radiative breaking of the
SU(2)$\times$U(1) symmetry.
The region below the bold line is ruled out by the experimental
bounds on the chargino mass $\mw1>45$ GeV and the
$Z^0$-boson decay widths to the neutralinos
$\sum_{i,j\neq 1}\Gamma(Z^0\rightarrow \chi_1\chi_i,
\chi_i\chi_j)< 5\times 10^{-2}$ MeV,
$\Gamma(Z^0\rightarrow \chi_1\chi_1)< 8.4$ MeV
\cite{susyexp}.
In the region below the lower dotted curve, the mass difference
between the lightest neutralino and the lighter stop becomes
larger than 5 GeV, which is ruled out by the recent direct stop
searches \cite{opal}.
In the region above the upper dotted curve, the lighter
stop becomes lighter than the lightest neutralino,
which is cosmologically disfavored.

     Within the presently allowed region, i.e. the region between
the two dotted curves
above the bold line, a large value of $R_C$ is predicted if
$\tb$ is not much larger than unity:
$R_C\ \gsim\ 2$ for $\tb\ \lsim\ 2$ and $R_C\ \gsim\ 1.5$
for $\tb\ \lsim\ 3$.
The chargino contribution and the $W$-boson contribution interfere
constructively, so that $R_C>1$.
If there exists a light stop, the new contribution
can be comparable to or even larger than the standard
model contribution.
The ratio $R_C$
increases as $\tb$ decreases, since a smaller value for
$v_2$ enhances the Yukawa couplings of the charginos to the
stops.

\begin{figure}
\begin{center}
\psfig{file=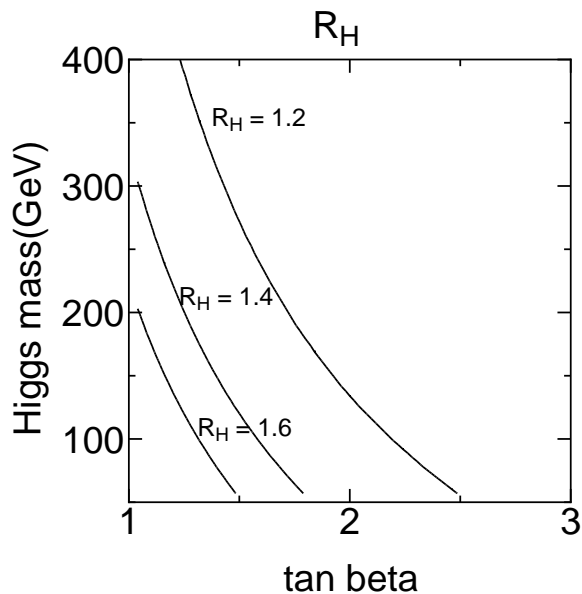,height=8cm}
\caption{The contours of the ratio $R_H$.}
\end{center}
\end{figure}
     The net effect of the MSSM is given by summing all the
contributions, for which the charged Higgs boson contribution
has to be evaluated.  In Fig. 2 we show the ratio
$R_H=(A_{tt}^W+A_{tt}^H)/A_{tt}^W$ as
contours in the ($\tb, \MCH$) plane, $\MCH$ being the
charged Higgs boson mass.
The charged Higgs boson contribution also interferes
constructively with the $W$-boson contribution.
Therefore, the value of $R$ in eq. (\ref{Ratio})
becomes larger than that of $R_C$ shown in Fig. 1.

     The value of $R$ affects the evaluation of the CKM
matrix elements through $x_d$ and $\epsilon$ \cite{bb2}.
Adopting the standard parametrization \cite{PDP}, the independent
parameters of the CKM matrix are represented by
three mixing angles $\theta_{12}, \theta_{13}, \theta_{23}$
and one $CP$-violating phase $\delta$.  Experimentally,
three mixing angles are determined by the processes for
which new contributions by the MSSM are negligible.
On the other hand, the $CP$-violating phase is
determined by $x_d$ or $\epsilon$, which depends on
the value of $R$.  Therefore, the value of $\delta$ in the
MSSM with $R>1$ is different from that in the SM of $R=1$.
Furthermore, $x_d$ and $\epsilon$ have to consistently give a
value of $\delta$ for a given $R$, which
constrains the values allowed for $\delta$ and $R$.

     The ranges of $R$ and $\cos\delta$ which are
consistent with the experimental results
for $x_d$ and $\epsilon$ have been
discussed in ref. \cite{bb2}.
Taking into account theoretical uncertainties
for hadronic matrix elements, it was shown
that the allowed range is $1\lsim R\lsim 2$ for
the experimental central values of $\sin\theta_{13}$
and $\sin\theta_{23}$.  Even if experimental uncertainties
for these mixing angles are taken into account, the value of
$R$ is at most 3.
It was also shown that
the measurements of $CP$ asymmetries of the $B$-meson
decays at B-factories could further constrain
the $R$ value.

     From these constraints on
the value of $R$, we can see that
most of the allowed region
for $\tb\lsim 2$ in Fig. 1
is inconsistent with \bbbar\ and \kkbar\ mixings.
Hence, the existence of a light stop with
$m_{{\tilde t}1}<M_Z/2$ can be
ruled out, if $\tb$ is around 2 or less.
In the allowed region for $\tb>2$, the value of $R$
becomes smaller than 2, which is not excluded from
the present data for \bbbar\ and \kkbar\ mixings.  However,
the allowed region for $\tb\lsim 3$, where $R>1.5$,
would be explored at B-factories in the near future.

     In conclusion, we have discussed the effects of
a light stop whose mass is less than $M_Z/2$ on \bbbar\
and \kkbar\ mixings.  The existence of such a light
stop gives large new contributions to those mixings
in sizable regions of the MSSM parameter space.
Examining these effects in the light of experimental
measurements of the $x_d$ and $\epsilon$ parameters,
we have shown that the light stop can be ruled out
if $\tb \lsim 2$.  For $\tb >2$, there are still
rooms for a light stop.  However, further constraints
will be obtained at B-factories.

\vspace{0.5cm}
     We thank S. Komamiya for correspondence.


\end{document}